\documentstyle[aps,prl,graphicx]{revtex}

\begin{document}
\draft

\twocolumn[\hsize\textwidth\columnwidth\hsize\csname@twocolumnfalse%
\endcsname

\title{Universality classes in anisotropic non-equilibrium growth models}

\author{Uwe C. T\"auber {}$^1$ and Erwin Frey {}$^{2,3}$}

\address{$^1$ Department of Physics, Virginia Tech, Blacksburg, 
  VA 24061-0435 \\
  $^2$ Hahn--Meitner Institut, Abteilung Theorie, D-14109 Berlin, 
  Germany \\ 
  $^3$ Freie Universit\"at Berlin, Fachbereich Physik, Arnimallee 14, 
  D-14195 Berlin, Germany}

\date{\today}

\maketitle

\begin{abstract}
  We study the effect of generic spatial anisotropies on the scaling
  behavior in the Kardar-Parisi-Zhang equation.  In contrast to its
  ``conserved'' variants, anisotropic perturbations are found to be
  relevant in $d > 2$ dimensions, leading to rich phenomena that
  include novel universality classes and the possibility of first-order 
  phase transitions and multicritical behavior.  These results question 
  the presumed scaling universality in the strong-coupling rough phase, 
  and shed further light on the connection with generalized driven 
  diffusive systems.
\end{abstract}

]

In non-equilibrium systems with {\em conserved} order parameter,
spatial anisotropies emerge as highly relevant perturbations, leading
to drastic changes in the universal scaling laws.  For example, in
driven diffusive systems (DDS) such as the driven Ising lattice gas
\cite{katz_lebowitz_spohn:83}, an external drive generates a
steady-state current that explicitly singles out one spatial
direction. This leads to characteristic singularities in the structure
factor already in the disordered phase, implies anisotropic scaling at
the phase transition, and generates rich ordered structures in the
low-temperature phase \cite{schmittmann_zia:95}.  Related anisotropic
scaling behavior even ensues near the critical point of an Ising or
Heisenberg system with conserved order parameter dynamics that is
driven out of equilibrium simply by imposing different Langevin noise
strengths in different spatial directions, thus violating detailed
balance \cite{schmittmann_zia:91,bassler_racz:94}.

Yet, one might expect that in models with a {\em non-conserved} order
parameter, the picture could rather look similar to the situation in
equilibrium systems, where anisotropies have much less dramatic
effects. In order to further explore this issue, we study the
Kardar--Parisi--Zhang (KPZ) model for kinetic roughening
\cite{kardar_parisi_zhang:86}, which has become another prototypical
model for generic scale invariance far from equilibrium
\cite{halpin-healy_zhang:95,krug:97}.  Its anisotropic generalization
reads
\begin{eqnarray}
  \label{eq:1}
  \partial_t h = \nu_\parallel \partial_\parallel^2 h + \nu_\perp
                 \partial_\perp^2 h + \frac{\lambda_\parallel}{2}
                 (\partial_\parallel h)^2 + \frac{\lambda_\perp}{2}
                 (\partial_\perp h)^2 + \eta  ,
\end{eqnarray}
with relaxation constants $\nu_{\parallel/\perp} > 0$, but no
restrictions on the signs of the non-linearities
$\lambda_{\parallel/\perp}$ that describe curvature-driven growth. We
denote the dimensionalities of the longitudinal and transverse sectors
with $d_\parallel$ and $d_\perp$, respectively, with $d_\parallel +
d_\perp = d$. $\eta$ is a stochastic driving force with Gaussian
correlations determined by the second moment $\langle \eta ({\bf
  x}',t') \eta({\bf x},t) \rangle = 2 D \delta({\bf x}-{\bf x}')
\delta (t-t')$. After a simple length rescaling one may of course
either choose $\nu_\parallel = \nu_\perp$ {\em or}
$|\lambda_\parallel| = |\lambda_\perp|$.

A previous study of this model restricted to two spatial dimensions
($d_\parallel = d_\perp = 1$) found the anisotropy to be irrelevant
for $\lambda_\parallel$ and $\lambda_\perp$ both positive 
\cite{wolf:91}. Interestingly, if the non-linear terms come with 
opposite sign, the anisotropy is still irrelevant, but in addition also 
the non-linearity itself scales to zero. In a related model, where one 
of the non-linear couplings vanishes, the relevance of the remaining
non-linearity was shown to depend on the values of $d_\parallel$ and
$d_\perp$ \cite{hwa:92}. It would therefore appear that anisotropies
play only a minor role in kinetic roughening phenomena: Only if we
render the signs of the non-linear couplings different, or set one of
them to zero, do we seem to obtain any change in the scaling behavior.
We shall, however, see that in any physical dimension larger than $2$,
the above anisotropies destroy isotropic scaling at long wavelengths
and in the long-time limit, and generate remarkably diverse behavior.
This supports earlier conjectures that perhaps the strong-coupling
scaling regime in the KPZ problem is {\em not} governed by universal
scaling exponents \cite{newman_swift:97}, but rather subtly depends on 
microscopic details. These are of course captured only rudimentarily 
in our simple anisotropies. For the original KPZ non-equilibrium 
growth or driven interface problem, this issue would appear to be 
mostly of academic interest; yet in light of a recent suggestion that 
the asymptotic scaling properties of interfaces pulled into unstable 
regions should be described by the $(d+1)$-dimensional rather than the 
$d$-dimensional KPZ equation \cite{tripathy_saarloos:00}, our findings 
may well become directly accessible to real experiments.

We have analyzed the non-linear Langevin equation (\ref{eq:1}) by
means of the dynamic renormalization group to one-loop order in the
perturbation expansion with respect to the non-linear couplings
$\lambda_{\parallel/\perp}$. The calculation is a straightforward
generalization of the one-loop treatment for the isotropic KPZ
equation or the equivalent noisy Burgers equation
\cite{forster_nelson_stephen:77,kardar_parisi_zhang:86}. We map the
Langevin equation to a dynamic functional, and proceed therefrom using
standard field-theoretic tools \cite{janssen:76}. The renormalization
constants that track the ultraviolet singularities in $d \geq 2$
dimensions are determined at a finite external wave vector $q = \mu$
(or equivalently, at non-zero external frequency)
\cite{frey_tauber:94}. The scaling behavior of the theory with respect
to this normalization scale is then encoded in the associated RG
equations, which are solved by the method of characteristics ($\mu \to
\mu l$).  The ensuing first-order differential flow equations define
the running couplings as functions of the flow parameter $l$ (or
momentum scale $\mu$). We have to distinguish between the cases where
both non-linearities have the same or opposite signs, $\lambda =
\lambda_\parallel = \pm \lambda_\perp$. The RG recursion relations $l
{\partial \alpha / \partial l} = \beta_\alpha (g,\gamma)$ for the 
{\em anisotropy ratio} $\gamma = \nu_\parallel / \nu_\perp$ and the 
{\em effective coupling} $ g = A_d \mu^{d-2} \lambda^2 D
\gamma^{d_\perp/2} / \nu_\parallel^3$, where $A_d = \Gamma
(2-\frac{d}{2}) / [2^{d+1} \pi^{d/2} d(d+2)]$ denotes a geometric
factor, are determined by the RG beta functions
\begin{eqnarray}
  \label{eq:2}
  \beta_g &=& g \left(d-2 + \zeta_D - 3 \zeta_{\nu_\parallel} +
              \frac{d_\perp}{2} \, \zeta_\gamma \right) \ , \\
  \label{eq:2a}  
  \beta_\gamma &=& \gamma \zeta_\gamma = \gamma (\zeta_{\nu_\parallel} -
              \zeta_{\nu_\perp}) \ , 
\end{eqnarray}
with the explicit one-loop flow functions
\begin{eqnarray}
  \label{eq:3}
  \zeta_D &=& -g \left[ d_\parallel (d_\parallel + 2) + d_\perp (d_\perp + 2) 
              \gamma^2 \pm 2 d_\parallel d_\perp \gamma \right] \ , \\
  \label{eq:3a}
  \zeta_{\nu_\parallel} &=& g \left[ d (d_\parallel \pm d_\perp \gamma) - 4
              \right] \ , \\
  \label{eq:3b}
  \zeta_{\nu_\perp} &=& g \gamma \left[ d (\pm d_\parallel + d_\perp \gamma) 
              - 4 \gamma \right] \  .
\end{eqnarray}
The recursion relations for the special case $\lambda_\perp =0$ are
readily obtained by setting $\gamma= 0$ in the expressions for these
zeta functions. The zeros of the beta functions (\ref{eq:2}),
(\ref{eq:2a}) yield the RG fixed points, describing scale-invariant
behavior. The universal infrared scaling exponents are then given by
the corresponding fixed-point values of the zeta functions
(\ref{eq:3})--(\ref{eq:3b}). Below the critical dimension $d_c=2$, the
theory is ultraviolet-finite, and the emerging RG fixed point is
infrared-stable. For $d \geq d_c$, the field theory remains
renormalizable as a consequence of the emergence of an infrared-{\em
  unstable} fixed point {\em and} the underlying infinitesimal tilt
invariance of the interface (Galilean invariance for the Burgers
equation) \cite{janssen_tauber_frey:99}. Technically,
renormalizability applies only in a systematic $\epsilon$ expansion
about the lower critical dimension for the roughening transition
($\epsilon = d - 2$). We shall nevertheless consider the theory also
at fixed $d = d_\parallel + d_\perp$, with either sector dimension or
$\Delta \equiv d_\parallel - d_\perp$ as parameters.

\begin{figure}[htbp]
  \centering
  \includegraphics[width=0.6\columnwidth,angle=90]{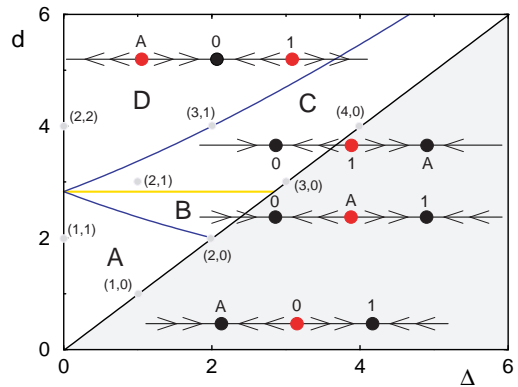}
  \caption{Stability diagrams for the fixed points of the anisotropy
    ratio $\gamma$ as functions of $d$ and $\Delta$; also indicated
    are $(d_\parallel,d_\perp)$ at some particulary interesting
    dimensions.}
  \label{fig:anisotropy_ratio}
\end{figure}

The first question to be addressed is whether the anisotropy is a
relevant perturbation at the isotropic KPZ fixed point. A previous
study established that the isotropic fixed point is stable in $2+1$
dimensions ($d_\parallel = d_\perp = 1$) \cite{wolf:91}.
However, our one-loop analysis reveals this to be far from true in
general. In fact, the RG flow equations, specifically 
Eq.~(\ref{eq:2a}) with (\ref{eq:3a}) and (\ref{eq:3b}), allow for 
{\em four} different scenarios, depending on the values of $d$ and 
$\Delta$. Without loss of generality we can restrict ourselves to 
$\Delta > 0$; for $\Delta < 0$ merely the roles of perpendicular and 
parallel components are switched. In an analysis at {\em fixed} $d$ 
and $\Delta$ we find the following regimes (depicted in 
Fig.~\ref{fig:anisotropy_ratio}):

In {\em regime A}, $d<-\Delta/2+\sqrt{(\Delta/2)^2+8}$, both the
isotropic fixed point $\gamma_1 = 1$ and the anisotropic fixed point
$\gamma_A = (4-d d_\parallel)/(d d_\perp -4)<0$ are stable with an
intermediate unstable uniaxial fixed point $\gamma_0 = 0$. Note (see
remark after flow equations) that a negative fixed point in the
anisotropy ratio does not mean that one of the relaxation constants
$\nu_\alpha$ becomes negative, but that we are at a fixed point where
the non-linearities have opposite sign. For the special case
$d_\parallel=d_\perp$ one obtains $\gamma_A =-1$, and thus Wolf's
earlier results~\cite{wolf:91} are recovered.
In {\em regime B}, $-\Delta/2+\sqrt{(\Delta/2)^2+8}<d<\sqrt{8}$, the
anisotropic fixed point becomes positive, $0<\gamma_A<1$, and switches
roles with the uniaxial fixed point $\gamma_0$.
In {\em regime C}, $\sqrt{8}<d<\Delta/2+\sqrt{(\Delta/2)^2+8}$, the
anisotropic fixed point becomes $\gamma_A > 1$ and again stable,
whereas the isotropic fixed point is now unstable.
Finally, in {\em regime D}, $d>\Delta/2+\sqrt{(\Delta/2)^2+8}$ , the
anisotropic fixed point assumes again a negative value. Yet here the
stability features of the fixed points are the converse of regime A.

Note that the stability boundaries for the various fixed points of the 
anisotropy ratio $\gamma$, as determined from Eq.~(\ref{eq:2a}), merely 
require the {\em existence} of a finite fixed point $0 < g^* < \infty$, 
and are independent of its actual value. Hence one may speculate that a 
topologically similar if not identical stability diagram applies for the 
critical fixed point $g_c$ even beyond our one-loop approximation, and 
might perhaps even extend to the ``strong-coupling'' scaling behavior 
in the rough phase.

\begin{figure}[htbp]
  \centering
  \includegraphics[width=0.5\columnwidth,angle=0]{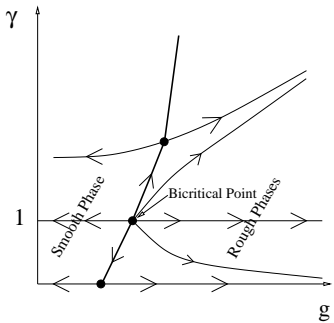}     
  \caption{Sketch of the projected renormalization group flows for the 
    anisotropic KPZ equation for scenario C.}
  \label{fig:RG_flow}
\end{figure}

The RG flows, projected onto the $\gamma$-$g$ plane, are illustrated
in Fig.~\ref{fig:RG_flow} for scenario C. There are four critical
fixed points, each one representing a different universality class
(as discussed below). In the neighborhood of the uniaxial and the
anisotropic fixed points, the flows along the critical surface are
towards each of them, since they each possess only one relevant
scaling variable, the effective coupling constant $g$. The fixed point
at $\gamma_1=1$ (describing the {\em isotropic} kinetic roughening
transition), located at the junction of two lines of critical points,
is an example of a {\em non-equilibrium bicritical point}. The RG flows
are quite reminiscent of a bicritical point in uniaxial magnetic
systems, with the high- and low-temperature phases of the magnetic
system corresponding to the smooth and rough phases of the kinetic
roughening problem. By analogy this suggests that there might actually
exist a {\em first-order} phase transition along the line $\gamma_1 = 1$ 
for $g>g_c$, separating two distinct non-equilibrium strong-coupling 
phases of different symmetry. Whereas the latter statement is somewhat
speculative, one can systematically study the critical behavior of the
new universality classes at the roughening transition.

If $\gamma \rightarrow 0$, i.e. $\nu_\perp \rightarrow \infty$, the
model reduces to the uniaxial KPZ equation with $\lambda_\perp = 0$.
This can be seen by direct inspection of the RG flow equation, but it
is also suggested by intuition since an infinitely large surface
tension $\nu_\perp$ will always dominate over any finite non-linearity. 
With ${\bf u} = \partial_\parallel h$ one realizes immediately that this
{\em uniaxial KPZ equation}\/ maps onto the {\em generalized driven 
diffusion equation} (GDDS)
\begin{equation}
  \label{eq:dds}
    \partial_t {\bf u} = (\nu_\parallel \partial_\parallel^2 + \nu_\perp
                 \partial_\perp^2) {\bf u} + \frac{\lambda_\parallel}{2}
                 \partial_\parallel {\bf u}^2 + \mbox{\boldmath $\zeta$} \ ,
\end{equation}
where ${\bf u}$ is a vector field with $d_\parallel$ components and
$\mbox{\boldmath $\zeta$}$ is a conserved noise with correlator
$\langle \zeta_i ({\bf x},t) \zeta_j ({\bf x}',t') \rangle = - 2 D
\delta_{ij} \nabla^2 \delta ({\bf x} - {\bf x'}) \delta (t - t')$.
Eq.~(\ref{eq:dds}) was originally introduced and analyzed to describe 
driven line liquids \cite{hwa:92}.

A crucial observation is that the nature of the fixed point of the
effective coupling constant $g$ changes drastically as a function of
$d$ and $\Delta$. This can be seen by inspecting the RG beta function
for $g$ in Eq.~(\ref{eq:2}). Note that the fixed-point value of the 
effective coupling constant diverges and changes sign at 
$\zeta_D = (3 - d_\perp/2) \zeta_{\nu_\parallel}$. In an $\epsilon$
expansion with respect to the critical dimension $d_c=2$ one then finds
two completely different regimes: For $\zeta_D < (3 - d_\perp/2)
\zeta_{\nu_\parallel}$, i.e., to leading order in $\epsilon$ for
$\Delta \geq 0.47 - 1.89 \epsilon$, there exists a critical fixed
point of order $\epsilon = d-2$ {\em above} two dimensions marking a
non-equilibrium phase transition from a smooth to a rough phase. In the 
converse case, the non-linearity is perturbatively irrelevant above 
$d_c=2$, and consequently the roughening transition disappears. In 
addition, {\em below} two dimensions one finds a stable fixed of order 
$\epsilon' = 2-d$ describing anomalous scaling behavior. In other words, 
there exists a critical value $\Delta_c \approx 0.47$ where $d_c=2$ 
changes its role from a {\em lower} to an {\em upper} critical 
dimension, and the elusive strong-coupling scenario is transformed into
a weak-coupling one.

These observations have several implications. First, owing to the GDDS
connection, it is possible to access the one-dimensional KPZ scaling 
($d_\parallel = 1$, $d_\perp = 0$) {\em perturbatively} by expanding 
around the upper critical dimension $d_c=2$ of the standard 
non-critical DDS model ($d_\parallel=1$, $d_\perp=1$). Second, a naive 
extrapolation of $\Delta = 0.47 - 1.89 \epsilon$ to $\epsilon = 1$ 
($d=3$) would suggest that DDS with $d_\parallel = 1$ and $d_\perp = 2$ 
might display a non-equilibrium phase transition. However, this is most 
likely an artifact of the $\epsilon$ expansion, and higher-order terms 
in the perturbation series will probably result in a boundary line for 
the critical fixed point that asymptotically approaches 
$d_\parallel = 1$ as $d_\perp \rightarrow \infty$ such that purely
mean-field behavior ensues for $d > 2$. Yet careful numerical 
simulations of the GDDS model that map out the $(d, \Delta)$-plane and 
determine the exact location of the boundary line, beyond the regime 
close to $d=2$ accessible in a perturbative RG approach, would
certainly be very desirable.

There are now several ways to calculate the exponents in an
$\epsilon$ expansion with respect to $d_c=2$. Keeping, e.g., 
$\Delta = d_\parallel - d_\perp$ fixed, we find $z_\parallel
= 2 + {2 \left( \Delta -2 \right) \,\epsilon }/{(\Delta^2 + 8 \Delta
  -4)}$ and $z_\perp =2$, which describes, depending on the value of
$\Delta$, the critical fixed point of the roughening transition or the
stable fixed point below $d_c$.  For example, extrapolating to 
$(d_\parallel,d_\perp)=(2,1)$ yields $z_\parallel \approx 1.6$. 
The actual value for $z$ found here is actually not crucial ---
the one-loop analysis should not be expected to be very accurate --- 
but the very fact that it differs from $z_c = 2$, the {\em exact} 
result for the roughening transition of the isotropic KPZ equation.
Furthermore, the RG analysis at the uniaxial fixed point shows no sign 
of an upper critical dimension, while it is known that $d=4$ is the 
upper critical dimension for the roughening transition of the 
isotropic KPZ model \cite{frey_tauber:94,lassig:95}.

Besides the uniaxial fixed points $\gamma_0=0$ and
$\gamma_\infty=\infty$ there exists also an anisotropic fixed point
with a {\em finite} anisotropy ratio $\gamma_A=(4-d d_\parallel)/(d
d_\perp -4)$. We find that the non-equilibrium dynamics described by
the anisotropic and isotropic fixed point are markedly different. At
first sight this may seem quite surprising since at both fixed points
the parallel and perpendicular surface tension term
$\nu_{\parallel,\perp}$ show {\em identical}\/ scaling behavior under
a RG transformation; the only difference resides in their amplitudes. 
The reason for this unusual behavior is that $\gamma = 1$ is a 
{\em non-generic} high-symmetry point in parameter space, allowing 
the isotropic KPZ problem to be mapped onto the statistical mechanics
of a directed polymer in a random medium. The high symmetry is also 
reflected in the mean-field value for the dynamic exponent $z_c=2$ at 
the {\em non}-trivial fixed point describing the roughening 
transition. In contrast, our RG analysis at the anisotropic fixed
point $\gamma_A$ yields a dynamic critical exponent {\em different} 
from $2$, to leading order in a $2+\epsilon$ expansion at fixed 
$\Delta$ given by $z = 2 + \epsilon \, (\Delta^2-4) / (10 \Delta^2 
-8)$. Similar to the GDDS equation, there exists a critical line 
$\Delta_c (d)$ such that for $|\Delta| < \Delta_c (d)$ there is no 
roughening transition above $d=2$. To leading order in $\epsilon$ we 
obtain $\Delta_c(d) = \pm (2/\sqrt{5} - 13 \epsilon / 5 \sqrt{5})$.  
Since $|\Delta_c(d)| < 1$ for $d>2$, a scenario with no roughening 
transition appears unlikely for any integer dimensions $d_\parallel$ 
and $d_\perp$. For $\Delta > \Delta_c (d)$ there is, as for the
uniaxial fixed point, one set of dimensions which is particularly
interesting, namely $(d_\parallel,d_\perp)=(2,1)$. Upon extrapolating 
the result obtained from our $\epsilon$ expansion we arrive at an
unphysically low critical value $z_c= 1/2 \pm O(\epsilon^2)$. 
Unfortunately, this leading-order result is not accurate enough to 
wager a reliable quantitative prediction for the actual value of the 
dynamic critical exponent. But, as before, the important conclusion 
to be drawn from the RG analysis is again that the dynamic critical 
exponent differs from the isotropic value $z_c=2$. In order to gain 
reliable quantitative estimates for the exponents of this novel
universality class, careful numerical simulations would be invaluable.

From the above analysis of the various universality classes and our 
understanding of the isotropic KPZ equation we derive the following
picture. With the exception of the particular case 
$d_\parallel=d_\perp$, the anisotropic fixed point always displays
strong-coupling behavior, i.e. we find a roughening transition above 
the lower critical dimension $d_c=2$ similar to the isotropic fixed 
point. However, the universality class of the roughening transition 
is markedly different from the isotropic KPZ model. In contrast to 
the anisotropic and the isotropic fixed points, the uniaxial fixed 
point $\gamma_0=0$ (GDDS) shows extended regions in the $(d,\Delta)$ 
plane with strong- and weak-coupling behavior, respectively. In 
particular, for $d_\parallel \leq d_\perp$ the roughening transition 
is entirely absent and we have anomalous scaling only below $d_c=2$.

Some specific examples serve to illustrate the implications of the 
above results. For $(d_\parallel,d_\perp)=(3,1)$ the isotropic fixed 
point $\gamma_1=1$ is unstable. If $\gamma < 1$, the RG flow tends 
towards the uniaxial fixed point. The corresponding GDDS universality 
class lacks the special symmetry of the isotropic KPZ equation 
implying that the dynamic critical exponent differs from $z_c =2$. 
If $\gamma > 1$, the RG trajectories flow towards $\gamma = \infty$. 
This fixed point lies again in the GDDS universality class, but with 
interchanged roles of $d_\parallel$ and $d_\perp$. Because of the 
symmetry of the model, the asymptotic behavior is equivalent to that
at the fixed point $\gamma_0 = 0$ in $(d_\parallel,d_\perp)=(1,3)$ 
space dimensions, where {\em no} roughening transition occurs above 
$d_c=2$. This implies not only that the isotropic KPZ equation is 
{\em unstable} in $d=4$ with respect to even the slightest amount of 
anisotropy, but more dramatically also that there exists {\em no}\/ 
phase transition and hence {\em no}\/ strong-coupling behavior at
all, provided the relaxation amplitudes satisfy $\nu_\parallel < 
\nu_\perp$ for $d_\parallel=1$. Similar critical end points are 
found throughout domain D of Fig.\ref{fig:anisotropy_ratio}.

Finally, we note that we have performed a similar one-loop RG analysis
for the ``conserved'' KPZ variants, with essentially an additional
Laplacian on the right-hand-side of Eq.~(\ref{eq:1}), both with
conserved \cite{sun_gao_grant:89} and non-conserved white noise 
\cite{wolf_villain:90}. Neither of these models displays a roughening 
transition. Rather, there is a single universal scaling regime with 
non-trivial exponents below the upper critical dimensions $d_c=4$ 
and $2$, respectively. It then turns out that anisotropic perturbations
do not matter crucially: The standard {\em isotropic} fixed point 
remains {\em stable} in both cases \cite{tauber_frey:01}. It would thus 
appear that the novel features reported above for the anisotropic KPZ 
equation are crucially connected with the very existence of a 
roughening transition, and thus with the scaling properties of a
strong-coupling rough phase.

It is a pleasure to acknowledge helpful discussions with H.K. Janssen,
B. Schmittmann, R.K.P. Zia, and T.J. Newman. U.C.T. is grateful for
support by the National Science Foundation (NSF Grant No. DMR-0075725)
and the Jeffress Memorial Trust (Grant No. J-594).  E.F. acknowledges
support by the Deutsche Forschungsgemeinschaft through a Heisenberg
fellowship (FR 850/3-1).

\end{document}